# Challenges in addressing student difficulties with time-development of two-state quantum systems using a multiple-choice question sequence in virtual and in-person classes

Peter Hu, Yangqiuting Li, and Chandralekha Singh
*Department of Physics and Astronomy, University of Pittsburgh, Pittsburgh, PA 15260*

**Abstract**

Research-validated clicker questions as instructional tools for formative assessment are relatively easy to implement and can provide effective scaffolding when developed and implemented in a sequence. We present findings from the implementation of a research-validated clicker question sequence (CQS) on student understanding of the time-development of two-state quantum systems. This study was conducted in an advanced undergraduate quantum mechanics course for two consecutive years in virtual and in-person classes. The effectiveness of the CQS discussed here in both modes of instruction was determined by evaluating students' performance after traditional lecture-based instruction and comparing it to their performance after engaging with the CQS.

## 1. Introduction

The time-evolution of a quantum state is an important concept in quantum mechanics and appears in many fields of active research, including the growing field of quantum information science. Since it draws on prerequisite knowledge of quantum states and the Hamiltonian of the system, the concept can be challenging for students to grasp on a first exposure. At the advanced undergraduate level, time-evolution of a quantum state is introduced with a time-independent Hamiltonian $\hat{H}$. The state as a function of time $t$ must satisfy the time-dependent Schrödinger equation $i\hbar \frac{d}{dt}|\chi(t)\rangle = \hat{H}|\chi(t)\rangle$, which for a time-independent Hamiltonian is equivalent to applying the time-evolution operator $e^{-\frac{i\hat{H}t}{\hbar}}$ to the initial state: $|\chi(t)\rangle = e^{-\frac{i\hat{H}t}{\hbar}}|\chi(t=0)\rangle$. When applied to the corresponding energy eigenstate $|\chi_n\rangle$, in accordance with the time-independent Schrödinger equation $\hat{H}|\chi_n\rangle = E_n|\chi_n\rangle$, the operator $\hat{H}$ yields the eigenvalue $E_n$. Therefore, $e^{-\frac{i\hat{H}t}{\hbar}}|\chi_n\rangle = e^{-\frac{iE_n t}{\hbar}}|\chi_n\rangle$. Since the energy eigenstates evolve in time via a trivial overall time-dependent phase factor, they are also known suitably as "stationary states."

As an example, a two-state system with Hamiltonian $\hat{H} = C\hat{S}_z$ with a dimensionally-appropriate constant $C$ will satisfy the time-independent Schrödinger equation for two stationary states: $\hat{H}|z\rangle = \frac{C\hbar}{2}|z\rangle$ and $\hat{H}|-z\rangle = -\frac{C\hbar}{2}|-z\rangle$, where $|z\rangle$ and $|-z\rangle$ represent the state with the z-component of its spin pointing "up" and "down," respectively. Any initial state can be expressed in the energy eigenbasis as $|\chi(0)\rangle = a|z\rangle + b|-z\rangle$, with $|a|^2 + |b|^2 = 1$. Applying the time-evolution operator to the two energy eigenstates replaces the Hamiltonian with the corresponding eigenvalue, so the state after time $t$ would be $|\chi(t)\rangle = e^{-\frac{i\hat{H}t}{\hbar}}|\chi(0)\rangle = ae^{-\frac{iCt}{2}}|z\rangle + be^{\frac{iCt}{2}}|-z\rangle$. If, however, the initial state is expressed in some other basis, one can obtain the state at time $t$ by first re-expressing the initial state as a superposition of energy eigenstates before introducing the time-dependent phase factors to each term.

To become proficient at determining a state at time $t$ given the initial state in some basis, students must be adept at several different tasks. These include being able to recognize whether a state is an eigenstate of the Hamiltonian, converting a state to the energy eigenbasis should the initial state be given in any other basis, and correctly applying the time-evolution operator.

Students also must recognize that different energy eigenstates generally correspond to different eigenvalues. The convergence of all these challenging concepts, as well as possible unfamiliarity with the meaning of the complex exponential itself, can place significant demands on students' cognitive resources. This may make other consequences of the Hamiltonian's central role in the evolution of a quantum state less obvious, e.g., the expectation value of any observable without explicit time-dependence does not depend on time in a stationary state.

Prior research suggests that students in quantum mechanics courses often struggle with many common difficulties [1–9], including with the basic formalism [10,11], notation [12], wavefunctions [13,14], the nature of probability [15], measurement [10,13,16,17], and transferring information to other contexts [7,18]. For such difficulties as those described, research-validated learning tools can effectively help students develop a robust knowledge structure [19–24]. For example, quantum interactive learning tutorials (QuILTs) have been developed, validated and implemented with encouraging results on many topics in quantum mechanics, including quantum measurement of physical observables [25–27], addition of angular momentum [28], perturbation theory and corrections to the energy spectrum of the hydrogen atom [29–32], systems of identical particles [13,33], quantum key distribution [34], Larmor precession [35], as well as the uncertainty principle and Mach-Zehnder interferometer [36,37]. Similarly, clicker questions, first popularized by Eric Mazur using his *Peer Instruction* method, are conceptual multiple-choice questions presented to a class for students to answer anonymously, individually first and again after discussion with peers, and with immediate feedback [38]. They have proven effective and are relatively easy to incorporate into a typical course, without the need to greatly restructure classroom activity or assignments [39]. When presented in sequences of validated questions, they can systematically help students with a particular theme that they may be struggling with. Previously, such clicker question sequences (CQS) have been successfully developed, validated and implemented on several key topics in quantum mechanics [40–44]. Critically, the time-evolution of a quantum state has also been identified as a common difficulty [10,45–48], but there have not been as many materials developed for time-evolution as for some other topics. Here we describe the development and validation of a CQS intended to help students learn time-evolution of two-state quantum systems.

## 2. Methodology

This research was carried out in accordance with the principles outlined in the University of Pittsburgh Institutional Review Board (IRB) ethical policy.

The CQS targets upper-level students in junior-/senior-level quantum mechanics courses. The data presented here are from implementation in a mandatory junior-/senior-level course at a large research university in the United States. To develop and validate this CQS, we studied the learning objectives and goals of the QuILT and CQS on similar topics that had previously been developed [35,36,43]. Taking inspiration from the pre- and post-tests validated alongside those QuILTs, we made adjustments to questions to specifically address the time-development of two-state systems. Additional inspiration came from questions from other sequences, including those focused on the time-development of quantum systems in the context of Larmor precession [35].

Additionally, we took advantage of much of the cognitive task analysis based on interviews, from both the expert and student perspectives, as well as the scaffolding that had been incorporated into the aforementioned QuILT. We focused on condensing this material to ensure that the CQS can be administered in the limited class time. To that end, we prioritized basic conceptual knowledge and specific consequences that students often find difficult, provided checkpoints at which the instructor should explain some broader themes related to the previous questions, and avoided burdensome calculations.

After we conceptualized the most important features that students should know about the time-development of quantum states, as well as a suitable organization and presentation of these features, we drafted, discussed, and iterated questions many times among ourselves to minimize unintended interpretations. We standardized terminologies and sentence constructions while simplifying them as much as possible to avoid causing cognitive overload for students. We also paid specific attention to the answer choices for each question. In some instances, we chose to pivot to a different learning goal after discussion revealed a more fundamental problem. Overall, we attempted to ensure that students would understand the questions unambiguously.

We aimed to address common stumbling blocks and emphasize key features that students may have missed in the large information content of a typical lecture. The thirteen questions in the CQS focused on the following four learning goals: identifying the basic properties of the energy eigenstates or stationary states (CQS 1.1-1.2) transforming from an initial state to its time-evolved state (CQS 2.1-2.5); expressing a state in the energy eigenbasis before applying the time-evolution operator (CQS 3.1-3.4); and calculating the time-dependence of the expectation values of various observables (CQS 4.1-4.2). We designed several questions specifically to address certain student difficulties that have previously been found. The questions used in the CQS are reproduced in appendix A.

The CQS was first implemented virtually in the fall of 2020 and was followed up by an in-person implementation in the fall of 2021, for which the major additional feature was Peer Instruction. We note that the instructors were different between the two classes. The CQS was administered in the virtual setting of 2020 as a Zoom poll while the instructor displayed the questions via the "Share Screen" function. For each question, the instructor displayed the results after all students had voted, before systematically discussing the validity of the options.

In a typical in-person classroom setting, students have easy access to one another to discuss their thinking in small groups. This, however, proved less feasible in the virtual instructional setting in 2020. When a majority of students selected an option that involved alternative conceptions, the instructor would give a hint and allow the students to vote again, or ask for volunteers to explain the reasoning behind their choices to serve as the hint. These cases, however, predominantly consisted of the student volunteers or instructor speaking to the whole class. As such, the Peer Instruction method as proposed by Mazur [38] was not implemented in full in the virtual classes, while it was achieved in the in-person classes in 2021.

To determine the effectiveness of the CQS in helping students overcome these common difficulties, we developed and validated a pre- and post-test, which had questions that were either taken directly from the CQS or on topics covered in the CQS. The post-test was a slightly modified version of the pre-test, containing changes such as a shift from eigenstates of $\hat{S}_x$ to eigenstates of $\hat{S}_y$, but otherwise remaining conceptually similar. In both virtual and in-person classes, students completed the pre-test immediately following traditional lecture-based instruction on the topic. After administration of the CQS, which took place over the course of three lecture sessions, students completed the post-test. For both, they were given a 25 min period at the end of the class session. Two researchers graded the pre-test and post-test and after discussion converged on a rubric, using which the inter-rater reliability was greater than 95%. Questions Q1, Q2, Q5 and Q6 provided students three possible answers from which to choose, and credit was awarded for correctly selecting or omitting each answer, for a total of up to three points. Questions Q3 and Q4 on the pre- and post-test were scored with two points split between answer and reasoning. On these questions, students were required to write their final answer without the operator $\hat{H}$ to receive full credit. The pre- and post-test questions are in appendix B.

## 3. Results and discussions

### *3.1. Lessons learned from virtual CQS administration*

The CQS in its final iteration (see appendix A) consisted of four subsections, each with questions focused on a specific learning goal. The two questions CQS 1.1-1.2 constituted a short primer and review that focused on identifying the basic properties of energy eigenstates, or stationary states. CQS 2.1-2.5 focused on transforming from an initial state to its time-evolved state. CQS 3.1-3.4 focused on expressing a state in the energy eigenbasis before applying the time-evolution operator. Finally, CQS 4.1-4.2 focused on unpacking and interpreting the time-dependence of the expectation values of various observables given an initial state.

Of the student difficulties intended to be addressed by the CQS, an item-by-item analysis indicated which ones students struggled with most. The lowest correctness percentage on any question was 27%, and the highest was 79%, rounded off for a class of N = 29 students. Some questions received a large majority of correct answers, but no question appeared to have an obviously correct answer selected by all students.

CQS 1.1 asked students to define a stationary state, and the results indicated that students did not fully understand the criteria for a stationary state. An eigenstate of *any* operator corresponding to a physical observable was a popular incorrect choice. Most students selected this in addition to an eigenstate of the Hamiltonian of the system, which is one of the correct choices for CQS 1.1, but not the only one. The prevalence of this difficulty is supported by prior research [11,48]. Furthermore, students initially answered that a superposition of stationary states must itself be a stationary state (CQS 1.2). That said, given a hint from the instructor noting that *any* state could be expressed as a superposition of stationary states and an opportunity to reconsider, more students selected the correct answer on the second round of polling.

CQS 1.2 asked students to select possible stationary states for a given system. Though 23% of the students had answered CQS 1.1 noting that an eigenstate of any operator is a stationary state, a higher percentage of students (32%) answered in CQS 1.2 that an eigenstate of $\hat{S}_x$ is a stationary state for a system with Hamiltonian $C\hat{S}_z$. In the second round of polling, the percentage of students who incorrectly noted that the eigenstate of $\hat{S}_x$ is a stationary state increased to 35%. It is possible that students got confused between different components of spin angular momentum, in light of previous comments made in class by the instructor that none of $\hat{S}_x$, $\hat{S}_y$, or $\hat{S}_z$ are inherently special.

The topic of CQS 2.2 was the time-development of a state that is not a stationary state. For this question, many students also initially incorrectly assumed that applying the unitary time-evolution operator results in a single time-dependent phase factor $e^{-\frac{iE_\pm t}{\hbar}}$ to the initial state, rather than using a different phase factor for each energy eigenstate [5,11]. Encouragingly, this tendency was observed less frequently in CQS 2.3 and CQS 2.5. In CQS 2.3, which asked about the time-development of an initial state $|y\rangle$, the answer choices are conceptually similar to those in CQS 2.2, while CQS 2.5 gave students several possible options from which to choose the correct expression for the time-development of a generic state. While the percentage of students choosing the correct answer remained relatively constant between 60-70%, this difficulty pertaining to attaching an overall phase $e^{-\frac{iE_\pm t}{\hbar}}$ to the state was selected with decreasing frequency. For CQS 2.5 specifically, the predominant incorrect answer omitted one of the correct choices rather than selecting the option consistent with this difficulty.

Many students were not comfortable with the properties of stationary states. In CQS 2.4, which had answer choices conceptually similar to those of CQS 2.2 but asked about an energy eigenstate, a sizable fraction believed that at least one of the choices was untrue, even though all of them were in fact correct statements. Some students appeared wary, after the preceding question, of multiplying the state by an overall time-dependent phase factor, even though the state in question is a stationary state. Others refrained from correctly answering that the outcome of a measurement of the observable $S_z$ is independent of time for this initial state. This could indicate difficulty in transferring a previous concept, the fact that a system in a stationary state remains in that state for all times $t$, to how this concept impacts measurements performed on this state at a time $t$. It is also possible that some students avoided choosing one or more of these options simply because they did not recognize that $|z\rangle$ is a stationary state in this problem. In CQS 2.5, some students were not comfortable taking the extra step from the correct expression $|\chi(t)\rangle = e^{\frac{-i\hat{H}t}{\hbar}}(a|z\rangle + b|-z\rangle)$ to the more explicit expression $|\chi(t)\rangle = ae^{\frac{-iE_+t}{\hbar}}|z\rangle + be^{\frac{-iE_-t}{\hbar}}|-z\rangle$, applying the time-evolution operator to each energy eigenstate in the superposition.

Additionally, students sometimes did not recognize when it was necessary to change the basis. This difficulty was first seen in CQS 2.3, then explicitly in CQS 3.2, in which the correct approach to find the time-evolved state for the system was to convert from the given basis to the $z$-basis (energy eigenbasis) before applying the time-evolution operator. In both questions, some students answered the questions by applying the time-evolution operator without changing the basis. For CQS 3.2, in which the given basis was the $x$-basis (which was not the energy eigenbasis), 45% of students selected "all of the above." This was the same proportion as those who selected the correct answer, which omitted a distractor choice involving the eigenstates of $\hat{S}_x$ written in the time-evolved state instead of the eigenstates of $\hat{S}_z$. Whether this was a careless mistake, or a genuine unawareness that the "energy eigenstates" mentioned in option III of CQS 3.2 were not in fact the basis states used in option II, was not clear. After the instructor allowed some students to voice their thoughts and provided hints, the class showed marked improvement in the subsequent CQS 3.3-3.4, both of which have initial states that required a change of basis before applying the time-evolution operator.

CQS 4.1-4.2 asked students to select the observables with time-independent expectation values for the given system, and the two questions illustrated the deep student difficulties with regard to expectation values of different observables in a given state. Each question invoked a specific consequence of Ehrenfest's Theorem: CQS 4.1 addressed the fact that an observable whose operator commutes with the Hamiltonian will have a time-independent expectation value, and CQS 4.2 covered the fact that the expectation values of all observables (with no explicit time-dependence) will be time-independent in a stationary state. For CQS 4.1, in which the system was in a non-stationary state, 50% of the students answered correctly, with the remaining students being split among the remaining answer choices. For CQS 4.2, in which the system was in a stationary state, students selected primarily between only two answer choices. The instructor gave students two chances to answer this question, but the distribution of selected answers remained very similar across both chances. 54% of students selected the correct answer on the first attempt, and 57% did so on the second. Most of the incorrect responses for CQS 4.2 selected only energy and $S_z$ while omitting $S_x$, whose corresponding operator does not commute with the Hamiltonian. These students had learned from the previous discussion that observables whose operators commute with the Hamiltonian have time-independent expectation values; however, they did not realize that the expectation values of all observables are time-independent because the initial state is a stationary state. Though it is possible that some students did not notice that the given state was a stationary state, this oversight may only be true for the first attempt, since the instructor pointed out before proceeding with the second round of polling that the state was indeed a stationary state.

Table 1 provides a summary of the student difficulties discussed in this section.

**Table 1.** The difficulties addressed by the CQS questions, which are found in appendix A.

| Difficulty | CQS # |
|---|---|
| 1. Any linear combination of stationary states is also a stationary state | 1.2 |
| 2. Difficulties with distinguishing between stationary states and eigenstates of *any* operator corresponding to an observable | 1.1, 2.1 |
| 3. Correctly recognizing that $\|\chi(t)\rangle = e^{-\frac{i\hat{H}t}{\hbar}}\|\chi(t=0)\rangle$, but not being able to write $\|\chi(t)\rangle$ with only the eigenvalues of $\hat{H}$ in place of the operator $\hat{H}$ | 2.5 |
| 4. Replacing the operator $\hat{H}$ with one eigenvalue, e.g., $E_{\pm}$, in the time-evolution operator $e^{-\frac{i\hat{H}t}{\hbar}}$, resulting in an overall phase | 2.2, 2.3 |
| 5. Not recognizing that a stationary state evolves via a trivial overall time-dependent phase factor | 2.4 |
| 6. Not recognizing that a stationary state will yield the same probabilities of measurement outcomes at time $t = 0$ and after a time $t$ for all observables (with no explicit time-dependence) | 2.4 |
| 7. Not focusing on the Hamiltonian and appropriately changing basis to find the state after time $t$ in terms of energies | 2.3, 3.1, 3.2 |
| 8. Mistakes in the process of changing basis | 3.3, 3.4 |
| 9. Time-evolution of expectation values: students unable to grasp that if $[\hat{H},\hat{Q}] = 0$, then $\frac{d}{dt}\langle\chi\|Q\|\chi\rangle = 0$ even if $\|\chi\rangle$ is not a stationary state | 4.1 |
| 10. Time-evolution of expectation values: students unable to grasp that in stationary states, none of the observables (with no explicit time-dependence) have time-dependent expectation values | 4.2 |

*3.2. Lessons learned from virtual and in-person pre- and post-test administration*

In contrast with its 2020 implementation as part of virtual instruction via Zoom, the CQS was administered in person in the fall of 2021 featuring fully-realized Peer Instruction [38]. The pre-test and post-test results (see questions in appendix B), as well as normalized gain [49] and effect sizes [50], are listed in tables 2 and 3 for virtual and in-person classes, respectively. We are encouraged that students in both virtual and in-person classes appear to have benefited from the CQS: across both years, from the pre-test to the post-test, effect sizes for questions ranged from 0.34 to over 1. A table that shows the concepts from the CQS on which students are evaluated in the pre- and post-test is provided as supplementary material (https://stacks.iop.org/EJP/43/025704/mmedia).

**Table 2.** Results of the 2020 virtual administration of the CQS via Zoom. Comparison of pre- and post-test scores, along with normalized gain [49] and effect size as measured by Cohen's *d* [50], for students who engaged with the CQS (N = 29).

| Q# | Pre-test mean | Post-test mean | Normalized gain | Cohen's *d* |
|---|---|---|---|---|
| 1 | 75% | 90% | 0.59 | 0.91 |
| 2 | 39% | 86% | 0.77 | 1.40 |
| 3 | 67% | 83% | 0.47 | 0.45 |
| 4 | 40% | 71% | 0.51 | 0.72 |
| 5 | 67% | 86% | 0.59 | 0.84 |
| 6 | 71% | 85% | 0.48 | 0.68 |

**Table 3.** Results of the 2021 in-person administration of the CQS. Comparison of pre- and post-test scores, along with normalized gain and effect size as measured by Cohen's *d*, for students who engaged with the CQS (N = 25).

| Q# | Pre-test mean | Post-test mean | Normalized gain | Cohen's d |
|---|---|---|---|---|
| 1 | 64% | 75% | 0.30 | 0.34 |
| 2 | 41% | 60% | 0.32 | 0.54 |
| 3 | 34% | 88% | 0.82 | 1.53 |
| 4 | 18% | 82% | 0.78 | 2.01 |
| 5 | 64% | 79% | 0.41 | 0.50 |
| 6 | 60% | 83% | 0.57 | 0.83 |

The overall average post-test scores after administration of the CQS improved to 83% in the virtual implementation and 77% in the in-person implementation. The somewhat lower average post-test scores for in-person classes in which Peer Instruction was incorporated fully may be for a variety of reasons, including the fact that students were different. However, one major reason could be that the pre- and post-tests in the virtual class were not proctored. Despite being told that these were closed-book and closed-notes, students in the virtual class who had their cameras off could have consulted with those resources, unlike in the in-person class. That said, students in the virtual class could have utilized such an advantage during both the pre-test and post-test, so the gain between the tests in both virtual and in-person classes is encouraging regardless, and many, though not all of the student difficulties addressed in the CQS were substantially reduced. Here we discuss each issue addressed, starting with cases in which student difficulties decreased significantly after CQS instruction, and then discussing one that was less successfully addressed.

*3.2.1. Difficulties with stationary states in general*

Questions Q1 and Q2 probed students' knowledge of stationary states, and among the highest normalized gains in 2020 were seen in these questions (see table 2). It appears that students knew reasonably well by the post-test that a linear combination of stationary states is not a stationary state, as demonstrated by performance on Q2. Students were able to transform from the given initial state in Q3 to the time-evolved state with a high success rate, which shows that they were better at applying the time-evolution operator on energy eigenstates than in the pre-test. By comparison, the post-test scores on these questions were lower in 2021, but student performance did still improve significantly from the pre-test.

*3.2.2. Confusion about generalized notation*

During the 2020 virtual administration of the CQS, a large number of students were not able write the expression $|\chi(t)\rangle = e^{-\frac{i\hat{H}t}{\hbar}}|\chi(t=0)\rangle$ so that it did not contain the Hamiltonian, when the initial state was given as $|\chi(t=0)\rangle = a|z\rangle + b|-z\rangle$ with $|a|^2 + |b|^2 = 1$ and the Hamiltonian as $\hat{H} = C\hat{S}_z$. However, most of the students did successfully rewrite in such a manner the expression on Q3 of the post-test, which gave a specific initial state $|\chi(t=0)\rangle = \frac{1}{\sqrt{5}}|x\rangle + \frac{2}{\sqrt{5}}|-x\rangle$ with Hamiltonian $\hat{H} = C\hat{S}_x$. Work from these students shows that they followed a procedure nearly identical to one that would be expected had the state instead been given as $a|x\rangle + b|-x\rangle$. The students also performed similarly well on Q3 in the 2021 in-person implementation, which is noteworthy given the much lower pre-test score, resulting in the largest observed normalized gain of 0.82 (see table 3).

*3.2.3. Difficulties with stationary states vs. eigenstates of any operator corresponding to observables other than energy*

Only the energy eigenstates of a system are the stationary states, but students often incorrectly associated the eigenstates of any generic operator with the stationary states. This difficulty could be exacerbated if students are shaky on the prerequisite linear algebra, without a clear grasp of what different eigenstates and operators mathematically or conceptually mean. Moreover, even if students are proficient with the linear algebra in the context of a math course, transferring that knowledge to the context of a quantum mechanics course can still be challenging. As illustrated by Q1 on the pre-test and post-test, students appeared to better distinguish between generic eigenstates and energy eigenstates after the CQS. As seen in table 2 for 2020 (table 3 for 2021), the correctness of this question improved from 75% (64%) to 90% (75%), with a normalized gain of 0.59 (0.30), which is at least partially due to improvement in this difficulty.

For post-test Q2, the given Hamiltonian commutes with $\hat{S}_z$, and the students more successfully avoided identifying the eigenstate of $\hat{S}_x$ (written in the *z*-basis) as a stationary state. Additionally on this question, students also better recognized that a superposition of stationary states is not a stationary state. 39% (41%) of students answered this question correctly on the pre-test, and 86% (60%) answered correctly on the post-test, with a normalized gain of 0.77 (0.32), indicating substantial improvement for 2020 and moderate improvement for 2021, again partially attributable to improvement in this difficulty. Similarly, students who correctly answered Q5 knew that the expectation values of only energy, and observables whose corresponding operators commute with the Hamiltonian, do not vary with time even in a non-stationary state.

*3.2.4. Replacing the operator $\hat{H}$ with one eigenvalue $E_n$*

During the virtual administration of CQS 2.2-2.5, many students selected an answer option that involved a single phase term containing energy instead of a sum of terms (distractor choice I in CQS 2.2 invokes this idea). This is at least partially due to students not being entirely comfortable with this concept, or not understanding the role of the Hamiltonian in the time-evolution of a general system. After CQS instruction, most students in both years correctly multiplied each energy eigenstate by a separate phase factor instead of a common phase factor for both Q3 and Q4. Final post-test performance was relatively strong for both questions, with normalized gains around 0.50 in 2020 and 0.80 in 2021 (see tables 2 and 3).

*3.2.5. Difficulties with change of basis*

Tables 2 and 3 show that, once students learned the importance of working in the appropriate basis, more were able to correctly answer Q4 on the post-test compared to the pre-test. In the 2020 virtual setting, post-test performance on Q4 is a bit lower than that on Q3, at 71% and 83%, respectively. These results generally carried over to the 2021 in-person setting, with students once again scoring slightly lower on Q4 (88%) than Q3 (82%); however, in both years, effect sizes were larger for Q4 than for Q3, so their improvement on Q4 was more uniform when compared to Q3. With the exception that in Q4 students needed to express the given state in the correct basis, the two questions were identical. Based upon class discussions, the lower performance on Q4 was mainly due to not recognizing the need for the basis change, or making a mistake in the basis change process.

*3.2.6. Expectation values of stationary states*

Improvement on Q6 appears similar in 2020 and 2021 according to tables 2 and 3; however, the number of students who selected correctly for all three choices (since Ehrenfest's theorem demands that they choose no observables' expectation values to depend on time in a stationary state) was relatively low even on the post-test. This seems to indicate a particularly resistant difficulty: students must recognize that probabilities of measurement outcomes in a stationary state do not depend on time, implying static expectation values. More scaffolding is evidently needed to help students learn this concept well. Moving forward, we would specifically include notes for instructors to encourage students to think of an expectation value as an average of a large number of measurements made on identically prepared systems, and how the probability of measuring different outcomes does not change in a stationary state. Students could also benefit from an additional discussion of Ehrenfest's theorem and the conditions that set $\frac{d}{dt}\langle Q \rangle = 0$ for an observable $Q$, giving them more tools with which to process these ideas and organize their knowledge [27].

### *3.3. Lessons learned from virtual and in-person class discussions*

A particular advantage of the CQS is that it provides opportunity for rich class discussions that can deepen student understanding. Despite the fact that the 2020 virtual format was not especially conducive to organic discussions among students, despite efforts made to that end, the instructor still engaged the class to the best of his ability. For CQS 1.2, which addressed the common difficulty that any superposition of stationary states is itself a stationary state, the correct answer was initially selected by only 29% of the class. Without revealing the answer, the instructor noted that the most general superposition of energy eigenstates could describe *any* possible state, and thus this option implied that every possible state is a stationary state. After this, on the second polling, the rate of correct answers rose to 48%, which is an improvement, though not a drastic one. In CQS 3.2, the Hamiltonian of the system is specified to commute with $\hat{S}_z$, but the initial state is given as a superposition of the eigenstates of $\hat{S}_x$. The instructor asked for two student volunteers to explain how the time-evolution of the state could not be expressed without the operator by remaining in that basis, and that option II, which replaced the Hamiltonian in the time-evolution operator with the energy eigenvalues without changing the initial basis, could not be correct. Finally, the topic for CQS 4.2 was the time-dependence of the expectation values of energy and components of spin in a stationary state, and despite the instructor's hint that the given state was an energy eigenstate (and thus a stationary state), the distribution of answers remained nearly identical both times the polling was opened to students. This points to the fact that this particular difficulty is robust, but it may also be attributable to the limitations of the CQS in a virtual format. While the students may not have been able to sufficiently parse the hint individually, it is likely that the performance would have improved in a typical classroom setting that would have allowed them to discuss the meaning of this hint and its consequences in small groups.

In the fall of 2021, the instructor reported lively, high-quality discussions for each clicker question. Though the post-tests do not appear to unambiguously reflect better overall student understanding as compared to 2020, the CQS was still successfully implemented as recommended by Mazur [38] and showed positive results. Students had abundant opportunities for more intimate discussions that often went for the whole time that they were allotted, and the comparison of post-test scores between the virtual and in-person implementations do not necessarily reflect the quality of discussion that occurred during class time.

Opportunities to hold an overall class discussion about salient concepts such as these after students have voted are very important, but ensuring that instructors hold such discussions when they are recommended can be a challenge especially because time is limited. We will continue to investigate ways to encourage such discussions via checkpoints between CQS questions, even in instances when the instructor may opt not to follow our suggestions verbatim.

## 4. Conclusions and summary

CQSs can be effective when implemented alongside traditional classroom lectures. We developed, validated, and found encouraging results from implementation of a CQS on the topic of time-development in two-state systems in virtual and in-person classes. Post-test scores improved for every question following the administration of the CQS in both settings. It is interesting to note that, from the pre-test to the post-test, students in 2020 showed overall more improvement in the multiple-answer questions Q1, Q2, Q5 and Q6, while the 2021 students improved much more dramatically in the free-response questions Q3 and Q4; this is evident in both the normalized gains and the effect sizes (see tables 2 and 3). As a whole, normalized gains appeared roughly uniform for the 2020 students, and noticeably larger for the open-ended questions compared to the multiple-choice ones in 2021. Meanwhile, effect sizes in 2020 were conventionally large to medium: most were over 0.70. In 2021, effect sizes were more polarized, with the multiple-choice questions giving values between 0.30-0.80, and the open-ended questions boasting remarkable effect sizes of 1.5-2.0.

Having considered these results, we have made some improvements for future administrations One of them is the inclusion of hints that ask students if a change of basis on a particular state is required before writing the time-evolved state without the Hamiltonian operator. Another is the inclusion of specific numbers as the expansion coefficients for states written in a particular basis, in addition to the general $a$ and $b$ with $|a|^2 + |b|^2 = 1$ that exclusively appeared in this CQS. Additionally, we have expanded the discussion on expectation values to address the physical interpretation of an expectation value as well as Ehrenfest's Theorem.

Regarding the somewhat surprising result that students performed equivalently or even better during virtual instruction, with inadequate Peer Instruction opportunities, we acknowledge that students may have been able to consult resources that they were not intended to during the virtually-administered pre- and post-tests. Even though they were told that it was a closed-book, closed-notes quiz, there was no way to adequately determine or enforce which resources students used when they had their cameras off. Even so, the students would have had access to those same resources during the pre-test, given after traditional lecture-based instruction, as well as the post-test, so the fact that post-test scores in the virtual class are significantly better is very encouraging. Moreover, we point out that students may have been stressed about the transition from virtual courses back to in-person learning, which may have affected performance for in-person class. Discussions with other instructors suggest that a greater number of students appear to be performing at a lower level in other physics courses after transition back to an in-person setting. In summary, the administration of the CQS in 2020 in a virtual learning context may have affected student performance compared to the in-person 2021 administration, but both classes benefited from the CQS. We note also that the results presented refer to the short- and mid-term retention of material, while analysis of long-term retention would require future studies and a longer lapse of time.

**Acknowledgements**
This research was carried out in accordance with the principles outlined in the University of Pittsburgh Institutional Review Board (IRB) ethical policy. We thank the NSF for award PHY-1806691.

# Appendix A

*The clicker questions in the sequence are reproduced below.*
- All of the questions refer to a two-state system.
- The notation $|\chi(0)\rangle$ indicates the state $|\chi\rangle$ at time $t = 0$, and $|\chi(t)\rangle$ indicates the state $|\chi\rangle$ at time $t$.
- The eigenstates of the operator $\hat{S}_z$ are written as $\{|z\rangle, |-z\rangle\}$, those of $\hat{S}_x$ are written as $\{|x\rangle, |-x\rangle\}$, and those of $\hat{S}_y$ are written as $\{|y\rangle, |-y\rangle\}$. The energy eigenvalues for the two-state system are denoted by $E_+$ and $E_-$.
- The Hamiltonian, and all other observables of interest, are assumed to have no explicit time-dependence.
- When a two-state system is written as $a|\chi_1\rangle + b|\chi_2\rangle$, it is assumed that $|a|^2 + |b|^2 = 1$, $|a| \neq 0$ and $|b| \neq 0$, and $a \neq b$.

## (CQS 1.1)
*Learning objective: Students are able to identify the properties of stationary states.*

Choose all of the following statements that are correct:
I. The term "stationary state" refers to an eigenstate of any operator that corresponds to a physical observable.
II. The term "stationary state" refers to an eigenstate of the Hamiltonian only.
III. A system in a stationary state will stay in this state for all times $t$ unless it is externally perturbed.

A. I only  
B. II only  
C. III only  
D. I and III only  
**E. II and III only**

## (CQS 1.2)
*Learning objective: Students are able to identify the properties of stationary states.*

Consider a system with a Hamiltonian $\hat{H} = C\hat{S}_z$. Choose all of the following initial states $|\chi(0)\rangle$ that are stationary states:
I. $|\chi(0)\rangle = |z\rangle$
II. $|\chi(0)\rangle = |x\rangle$
III. $|\chi(0)\rangle = a|z\rangle + b|-z\rangle$, because $|z\rangle$ and $|-z\rangle$ are both stationary states.

A. All of the above  
**B. I only**  
C. I and II only  
D. I and III only  
E. None of the above

*Class discussion for CQS 1.1-1.2*
- *The eigenstates of the Hamiltonian are known as stationary states and have trivial time evolution, in which the state at time t depends upon an overall time-dependent phase factor, e.g., $|\chi(t)\rangle = e^{-\frac{i\hat{H}t}{\hbar}}|\chi(0)\rangle = e^{-\frac{iE_+t}{\hbar}}|z\rangle$ if $|\chi(0)\rangle = |z\rangle$ is a stationary state.*
- *A superposition of stationary states is not a stationary state if there is no degeneracy.*

## (CQS 2.1)
*Learning objective: Students are able to make a distinction between eigenstates of the Hamiltonian (and operators that commute with the Hamiltonian) and eigenstates of any other operator.*

Note: Any observable (which, for generality, we'll call $Q$) has a corresponding Hermitian operator $\hat{Q}$, which has a complete set of eigenstates $|q_+\rangle$ and $|q_-\rangle$ with eigenvalues $q_+$ and $q_-$ respectively. Assume that $\hat{Q}$ does not commute with the Hamiltonian.

Choose all of the following that are correct about the time-development of a generic state $|\chi(0)\rangle$:
I. The time evolution of the state is governed by the Hamiltonian operator for that system.
II. If $|\chi(0)\rangle$ is an energy eigenstate (also known as a stationary state), it will remain an energy eigenstate after a time $t$.
III. If $|\chi(0)\rangle$ is an eigenstate of $\hat{Q}$, it will remain an eigenstate of $\hat{Q}$ after a time $t$.

A. All of the above     B. I only     C. II only
**D. I and II only**     E. II and III only

**(CQS 2.2)**
*Learning objective: Students are able to describe the effect of the time-evolution operator when applied to a generic non-stationary state.*

Choose all of the following that are correct about the time-development of a state $|\chi(0)\rangle$, which in this case is **not a stationary state**:

I. $|\chi(t)\rangle = e^{\frac{-iE_+ t}{\hbar}} |\chi(0)\rangle$ or $e^{\frac{-iE_- t}{\hbar}} |\chi(0)\rangle$, where $E_+$ and $E_-$ are eigenvalues of $\hat{H}$.
II. $|\chi(t)\rangle = e^{\frac{-i\hat{H}t}{\hbar}} |\chi(0)\rangle$, where $\hat{H}$ is the Hamiltonian of the system.
III. The probability of measuring $q_+$ and $q_-$ in the state $|\chi(t)\rangle$ will be the same, regardless of the time $t$ when the measurement is performed. [51]

**A. II only**     B. III only     C. I and II only
D. II and III only     E. None of the above

**(CQS 2.3)**
*Learning objective: Students are able to describe the effect of the time-evolution operator $e^{\frac{-i\hat{H}t}{\hbar}}$ when applied to a concrete example of an initial state which is a non-stationary state not written as a superposition.*

Consider a system with a Hamiltonian $\hat{H} = C\hat{S}_z$. Choose all of the following that are correct at time $t$ about the system if the initial state is $|\chi(0)\rangle = |y\rangle$:

I. $|\chi(t)\rangle = e^{\frac{-iE_+ t}{\hbar}} |\chi(0)\rangle$
II. $|\chi(t)\rangle = e^{\frac{-i\hat{H}t}{\hbar}} |\chi(0)\rangle$
III. If the observable $S_y$ is measured, the probability of obtaining $\frac{\hbar}{2}$ is 1, *regardless of time $t$ when the measurement is performed.*

A. All of the above     B. I only     **C. II only**
D. I and II only     E. II and III only

**(CQS 2.4)**
*Learning objective: Students are able to describe that the effect of the time-evolution operator when applied to a stationary state is to evolve the state by an overall time-dependent phase factor (i.e., the state remains a stationary state).*

Consider a system with a Hamiltonian $\hat{H} = C\hat{S}_z$. Choose all of the following that are correct at time $t$ about the system if the initial state is $|\chi(0)\rangle = |z\rangle$:

I. $|\chi(t)\rangle = e^{\frac{-iE_+ t}{\hbar}} |\chi(0)\rangle$
II. $|\chi(t)\rangle = e^{\frac{-i\hat{H}t}{\hbar}} |\chi(0)\rangle$
III. If the observable $S_z$ is measured, the probability of obtaining $\frac{\hbar}{2}$ is 1, *regardless of time $t$ when the measurement is performed.*

A. I only  B. I and II only  C. I and III only
D. II and III only  **E. All of the above**

**(CQS 2.5)**
*Learning objective: Students are able to describe the effect of the time-evolution operator when applied to a quantum state written as a superposition of the eigenstates of $\hat{H}$.*

Consider a system with a Hamiltonian $\hat{H} = C\hat{S}_z$ in the state $|\chi(0)\rangle = a|z\rangle + b|-z\rangle$. Choose all of the following expressions that are correct for the state $|\chi(t)\rangle$ at time $t$:

I. $|\chi(t)\rangle = e^{\frac{-iE_+t}{\hbar}}(a|z\rangle + b|-z\rangle)$

II. $|\chi(t)\rangle = e^{\frac{-i\hat{H}t}{\hbar}}(a|z\rangle + b|-z\rangle)$

III. $|\chi(t)\rangle = ae^{\frac{-iE_+t}{\hbar}}|z\rangle + be^{\frac{-iE_-t}{\hbar}}|-z\rangle$

A. I only  B. II only  C. III only
D. I and II only  **E. II and III only**

*Class discussion for CQS 2.1-2.5*
- The previous 5 questions may be asked consecutively, with a discussion at the end. They involve several different situations, from general cases to concrete examples.
- $|\chi(t)\rangle = e^{\frac{-i\hat{H}t}{\hbar}}|\chi(0)\rangle$ *is always true, but* $|\chi(t)\rangle = e^{\frac{-iE_+t}{\hbar}}|\chi(0)\rangle$ *(an overall phase factor) is true only if* $|\chi(0)\rangle$ *is the eigenstate of the Hamiltonian corresponding to the energy* $|E_+\rangle$ *or* $|E_-\rangle$.

**(CQS 3.1)**
*Learning objective: Students are able to describe how the completeness relation (spectral decomposition of the identity operator) can be used to write a generic state in a basis consisting of energy eigenstates. Other learning objective are similar to those in CQS 2.5.*

Consider a system with a Hamiltonian $\hat{H} = C\hat{S}_z$. Choose all the correct statements about a system in the state $|\chi(0)\rangle$:

I. $|\chi(0)\rangle = (|z\rangle\langle z| + |-z\rangle\langle -z|)|\chi(0)\rangle$, where $|z\rangle\langle z| + |-z\rangle\langle -z| = \hat{I}$

II. $|\chi(0)\rangle = C_1|z\rangle + C_2|-z\rangle$, where $C_1 = \langle z|\chi(0)\rangle$ and $C_2 = \langle -z|\chi(0)\rangle$

III. $|\chi(t)\rangle = e^{\frac{-iE_+t}{\hbar}} C_1|z\rangle + e^{\frac{-iE_-t}{\hbar}} C_2|-z\rangle$

A. I only  B. II only  C. I and II only
D. II and III only  **E. All of the above**

**(CQS 3.2)**
*Learning objective: Students are able to identify that the state first must be written in the energy eigenbasis before the time-evolution operator can be used to write the state after time $t$ in the energy eigenbasis without the operator $\hat{H}$.*

Consider a system with a Hamiltonian $\hat{H} = C\hat{S}_z$. Choose all of the following that are correct about a system in the state $|\chi(0)\rangle = a|x\rangle + b|-x\rangle$.

I. $|\chi(t)\rangle = e^{\frac{-i\hat{H}t}{\hbar}}|\chi(0)\rangle$

II. $|\chi(t)\rangle = ae^{\frac{-iE_+t}{\hbar}}|x\rangle + be^{\frac{-iE_-t}{\hbar}}|-x\rangle$

III. To find $|\chi(t)\rangle$, we can write $|\chi(0)\rangle$ as a linear superposition of energy eigenstates, and then attach a time-dependent phase factor with the appropriate energy to each term.

A. I only  B. III only  C. I and II only
**D. I and III only**  E. All of the above

**(CQS 3.3)**

*Learning objective: Students are able to identify that the state must be in the energy eigenbasis before the time-evolution operator can be applied in order to obtain an explicit expression for the time-dependent state. When the given initial state is not written in terms of the energy eigenbasis, they are able to write the time-evolved state in this basis.*

Consider a system with a Hamiltonian $\hat{H} = C\hat{S}_z$. Choose the correct expression for the time-evolved state $|\chi(t)\rangle$ given an initial state $|\chi(0)\rangle = a|x\rangle + b|-x\rangle$.

A. $|\chi(t)\rangle = a e^{\frac{-iE_+ t}{\hbar}}|x\rangle + b e^{\frac{-iE_- t}{\hbar}}|-x\rangle$

B. $|\chi(t)\rangle = \frac{a}{\sqrt{2}} e^{\frac{-iE_+ t}{\hbar}}|z\rangle + \frac{b}{\sqrt{2}} e^{\frac{-iE_- t}{\hbar}}|-z\rangle$

C. $|\chi(t)\rangle = \frac{a+b}{\sqrt{2}} e^{\frac{-iE_+ t}{\hbar}}|x\rangle + \frac{a-b}{\sqrt{2}} e^{\frac{-iE_- t}{\hbar}}|-x\rangle$

D. $|\chi(t)\rangle = \frac{a+b}{\sqrt{2}} e^{\frac{-iE_+ t}{\hbar}}|z\rangle + \frac{a-b}{\sqrt{2}} e^{\frac{-iE_- t}{\hbar}}|-z\rangle$

E. None of the above

**(CQS 3.4)**

*Learning objective: Students are able to identify that the state must be in the energy eigenbasis before the time-evolution operator can be applied in order to obtain an explicit expression for the time-dependent state.. Here, $\hat{H} \propto \hat{S}_x$, so a transformation from the z-basis to the x-basis is required.*

Consider a system with a Hamiltonian $\hat{H} = C\hat{S}_x$. Choose all of the following that are correct about the time-development of the state $|\chi(0)\rangle = \frac{1}{\sqrt{2}}|z\rangle + \frac{1}{\sqrt{2}}|-z\rangle$:

I. $|\chi(t)\rangle = \frac{1}{\sqrt{2}} e^{\frac{-iE_+ t}{\hbar}}|z\rangle + \frac{1}{\sqrt{2}} e^{\frac{-iE_- t}{\hbar}}|-z\rangle$

II. $|\chi(t)\rangle = \frac{1}{\sqrt{2}} e^{\frac{-iE_+ t}{\hbar}}|x\rangle + \frac{1}{\sqrt{2}} e^{\frac{-iE_- t}{\hbar}}|-x\rangle$

III. $|\chi(t)\rangle = e^{\frac{-iE_+ t}{\hbar}}|x\rangle$

A. I only  
B. II only  
**C. III only**  
D. I and III only  
E. None of the above

*Class discussion for CQS 3.1-3.4*

- The instructor can choose to give the following relations to the students for CQS 3.3-3.4.

$\langle z|x\rangle = \frac{1}{\sqrt{2}}$  $\langle -z|x\rangle = \frac{1}{\sqrt{2}}$

$\langle z|-x\rangle = \frac{1}{\sqrt{2}}$  $\langle -z|-x\rangle = -\frac{1}{\sqrt{2}}$

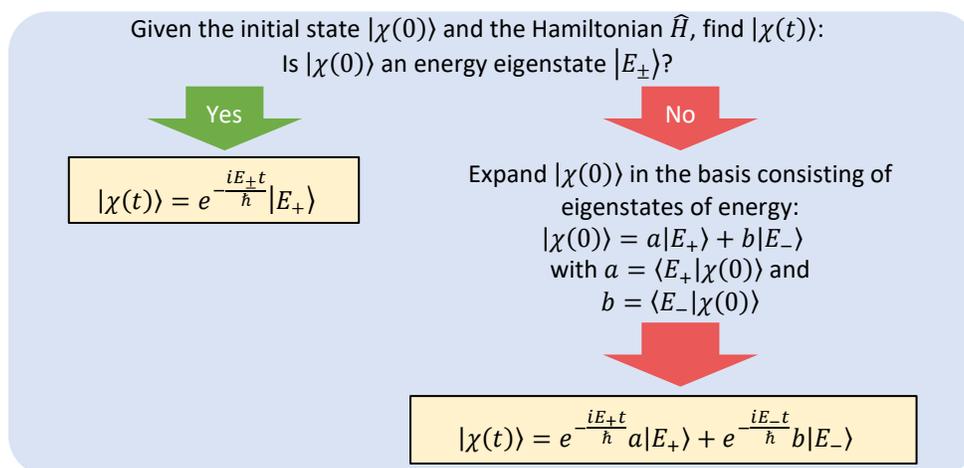

**Figure 1.** A flowchart detailing the process of evolving a given initial state in time, provided to students as part of class discussion.

**(CQS 3.5)**
*Learning objective: Students are able to identify that the expectation values of energy and all observables whose corresponding operators commute with the Hamiltonian are independent of time, regardless of the initial state.*

Consider a system with a Hamiltonian $\hat{H} = C\hat{S}_z$. If the system is in the state $|\chi(0)\rangle = a|z\rangle + b|-z\rangle$, choose all of the following observables whose expectation values are time-independent:

I. Energy
II. $S_z$
III. $S_x$

A. None of the above
B. I only
**C. I and II only**
D. II and III only
E. All of the above

**(CQS 3.6)**
*Learning objective: Students are able to identify that the expectation values of all observables for a stationary state are independent of time.*

Consider a system with a Hamiltonian $\hat{H} = C\hat{S}_z$. If the system is in the state $|\chi(0)\rangle = |z\rangle$, choose all of the following observables whose expectation values are time-independent:

I. Energy
II. $S_z$
III. $S_x$

A. None of the above
B. I only
C. I and II only
D. II and III only
**E. All of the above**

*Class discussion for CQS 4.1-4.2*
- *The expectation value of energy is always time-independent regardless of the state (so long as $\hat{H}$ does not explicitly depend on time).*
- *$S_z$ is special in this instance, because it is an observable whose corresponding operator commutes with the Hamiltonian. Its expectation value is also time-independent, regardless of the state.*
- *Expectation values of other observables such as $\langle S_x \rangle$ and $\langle S_y \rangle$ are time-dependent in CQS 4.1. But if the Hamiltonian were proportional to $\hat{S}_x$, $\langle S_x \rangle$ would be time-independent regardless of the state, while $\langle S_z \rangle$ would be time-dependent.*

# Appendix B

*The pre- and post-test questions are reproduced below. The same information provided at the beginning of Appendix A applies to the questions in Appendix B. Students were also given the following information:*

$$\langle x|z\rangle = \frac{1}{\sqrt{2}} \quad \langle -x|z\rangle = \frac{1}{\sqrt{2}}$$
$$\langle x|-z\rangle = \frac{1}{\sqrt{2}} \quad \langle -x|-z\rangle = -\frac{1}{\sqrt{2}}$$

<u>Note:</u> Any observable (which, for generality, we'll call $Q$) has a corresponding Hermitian operator $\hat{Q}$, which has a complete set of eigenstates $|q_+\rangle$ and $|q_-\rangle$ with eigenvalues $q_+$ and $q_-$ respectively. Assume that $\hat{Q}$ does not commute with the Hamiltonian.

**(Q1)** Choose all of the following that are correct about the time-development of a state $|\chi(0)\rangle$, which in this case is **not a stationary state**:

I. $|\chi(t)\rangle = e^{\frac{-iE_+t}{\hbar}}|\chi(0)\rangle$ or $e^{\frac{-iE_-t}{\hbar}}|\chi(0)\rangle$, where $E_+$ and $E_-$ are eigenvalues of $\hat{H}$.

II. $|\chi(t)\rangle = e^{\frac{-i\hat{H}t}{\hbar}}|\chi(0)\rangle$, where $\hat{H}$ is the Hamiltonian of the system.

III. The probability of measuring $q_+$ and $q_-$ in the state $|\chi(t)\rangle$ will be the same, regardless of the time $t$ when the measurement is performed. [51]

**(Q2)** Consider a system with a Hamiltonian $\hat{H} = C\hat{S}_z$, where $C$ is an appropriate constant. Choose all of the following that are stationary states for this system:
I. $\frac{1}{\sqrt{2}}|z\rangle + \frac{1}{\sqrt{2}}|-z\rangle$
II. $a|z\rangle + b|-z\rangle$
III. $a|x\rangle + b|-x\rangle$, where $a \neq b$

**(Q3)** Consider a system with a Hamiltonian $\hat{H} = C\hat{S}_x$ (note: $\hat{H} \neq C\hat{S}_z$), where $C$ is an appropriate constant. For a system in the initial state $|\chi(t=0)\rangle = \frac{1}{\sqrt{5}}|x\rangle + \frac{2}{\sqrt{5}}|-x\rangle$, what is $|\chi(t)\rangle$? Show your work.

**(Q4)** Consider a system with a Hamiltonian $\hat{H} = C\hat{S}_x$, where $C$ is an appropriate constant. For a system in the initial state $|\chi(0)\rangle = \frac{1}{\sqrt{5}}|z\rangle + \frac{2}{\sqrt{5}}|-z\rangle$, what is $|\chi(t)\rangle$? Show your work.

**(Q5)** Consider a system with a Hamiltonian $\hat{H} = C\hat{S}_z$, where $C$ is an appropriate constant. For a system in the initial state $|\chi(0)\rangle = a|z\rangle + b|-z\rangle$, choose all of the following observables whose expectation values are time-independent:
I. Energy
II. $S_y$
III. $S_z$

**(Q6)** Consider a system with a Hamiltonian $\hat{H} = C\hat{S}_z$, where $C$ is an appropriate constant. For a system in the state $|\chi(0)\rangle = |z\rangle$, choose all of the following observables whose expectation values are time-independent:
I. Energy
II. $S_y$
III. $S_z$